\documentstyle[natbib,epsfig,psfig]{elsart}
\begin{document}
\begin{frontmatter}

\title{
\hfill \vbox{ \hbox{hep-ph/0001129}  
\hbox{FTUV/00-04} 
\hbox{IFIC/00-04}}\\
Neutrino Masses and Mixing one Decade from Now}
\author{M. C. Gonzalez-Garcia and C. Pe\~na-Garay}
\address{Dept. de F\'{\i}sica Te\`orica \\
Inst. de F\'{\i}sica Corpuscular \\
C.S.I.C. - Univ. de Val\`encia, Spain}
\begin{abstract}
We review the status of neutrino masses and mixings
in the light of the solar and atmospheric neutrino data.
The result from the LSND experiment is also considered.
We discuss the present knowledge and the expected sensitivity
to the neutrino mixing parameters in the 
simplest schemes proposed to 
reconcile these data some of which include a light sterile neutrino 
in addition to the three standard ones. 
      \footnote{
        Talk given at the ICFA/ECFA Workshop "Neutrino Factories based 
        on Muon Storage Rings, $\nu$-FACT99", Lyon, July 1999.}
\end{abstract}
\end{frontmatter}

\section{Indications for Neutrino Mass: Two--Neutrino Analysis}

Neutrinos are the only massless fermions predicted by the Standard Model. 
This seemed to be a reasonable assumption as none of the 
laboratory experiments designed to measure
the neutrino mass have found any positive
evidence for a non-zero neutrino mass. 
However, the confidence on the masslessness of the neutrino is now
under question due to the important results of underground experiments,
starting by the geochemical experiments of Davis and collaborators till 
the more recent Gallex, Sage, Kamiokande and Super--Kamiokande (SK)
experiments \cite{solarexp,atmexp,sk99}. 
Altogether they provide solid evidence for the existence of anomalies
in the solar and the atmospheric neutrino fluxes which could be 
explained by the hypothesis of neutrino oscillations which requires
the presence of neutrino masses and mixings.   
Particularly relevant has been the recent confirmation by the
SK collaboration \cite{sk99} of the atmospheric
neutrino zenith-angle-dependent deficit which strongly indicates towards
the existence of $\nu_\mu$ conversion. Together with these results 
there is also the indication for
neutrino oscillations in the $\bar\nu_\mu \rightarrow \bar\nu_e$ channel 
by the LSND experiment~\cite{lsnd}. 

We first review our present knowledge of the present experimental 
status for the different evidences and present the results of the
different analysis in the framework of two--neutrino oscillations. 

\subsection{Solar Neutrinos}
\label{solar}
At the moment, evidence for a solar neutrino deficit comes from five
experiments \cite{solarexp}: Homestake, Kamiokande, SK 
and the radiochemical Gallex and Sage experiments.
The most recent data on the rates can be summarized as: 
\begin{eqnarray}
\mbox{Clorine} & & 2.56 \pm 0.23\;\; \mbox{SNU} \nonumber \\ 
\mbox{Gallex and Sage} &  & 72.2 \pm
5.6 \;\;\mbox{SNU} \nonumber \\
\mbox{Super--Kamiokande} & & 
(2.45 \pm 0.08) \times 10^6\;\;\mbox{cm$^{-2}$s$^{-1}$} \nonumber
\end{eqnarray}
Super-Kamiokande has also measured the 
the time dependence of the event rates during the day and night, as
well as a measurement of the recoil electron energy spectrum and has 
also presented
preliminary results on the seasonal variation of the neutrino event
rates, an issue which will become important in discriminating the MSW
scenario from the possibility of neutrino oscillations in
vacuum.

The different experiments are sensitive to different parts of the 
energy spectrum of solar neutrinos and
putting all these results together seems to 
indicate that the solution to the problem is not astrophysical but must
concern the neutrino properties. Moreover, non-standard astrophysical 
solutions are strongly constrained
by helioseismology studies \cite{Bahcall98}. Within the
standard solar model approach, the theoretical predictions clearly lie
far from the best-fit solution what leads
us to conclude that new particle physics is the only
way to account for the data.
\begin{figure}
\centerline{\protect\hbox{\psfig{file=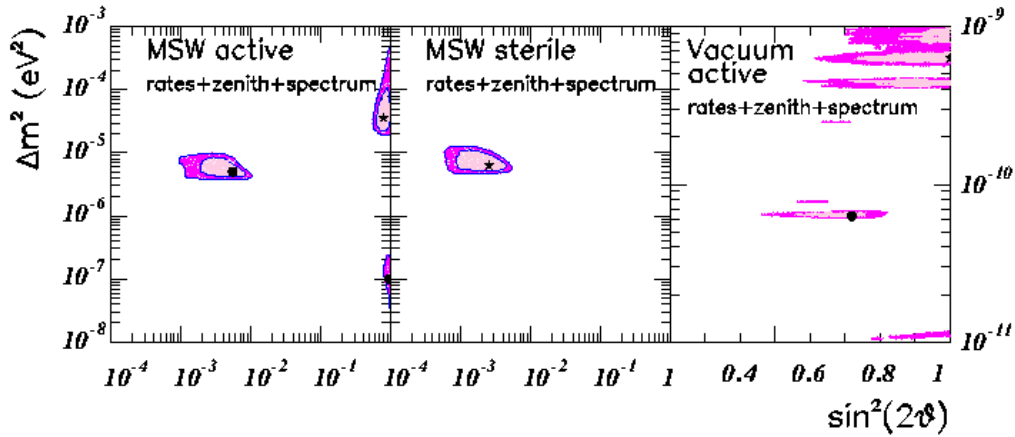,width=\textwidth}}}
\vglue -0.5cm
\caption{Presently allowed solar neutrino parameters for two-neutrino
oscillations by the global analysis from Ref.~\protect\cite{oursun,ourfour}. 
The plotted regions are 90\% (lighter) and 99\% CL (darker).}
\label{solar2}
\end{figure}

The standard explanation for this deficit
is the oscillation of $\nu_e$ to another neutrino species either active
or sterile. In Fig.~\ref{solar2} we show the allowed two-neutrino oscillation 
regions obtained in our updated global analysis of the solar neutrino data 
\cite{oursun,ourfour} for both MSW \cite{msw}  
as well as vacuum oscillations \cite{vacuum} into active 
or sterile neutrinos. These results indicate that for oscillations into 
active neutrinos there are four possible solutions for the parameters:
\begin{itemize}
\item [$\bullet$] vacuum (also called ``just so'') oscillations with 
$\Delta m^2_{ei}=(0.5$--$8)\times 10^{-10}$  eV$^2$ and 
$\sin^2(2\theta)=0.5$--$1$ 
\item[$\bullet$] non-adiabatic-matter-enhanced oscillations (SMA) via 
the MSW mechanism  
with $ \Delta m^2_{ei}=(0.4$--$1)\times 10^{-5}$ eV$^2$ and 
$\sin^2(2\theta)=(1$--$10)\times 10^{-3} $, and 
\item[$\bullet$] large mixing (LMA) via the MSW mechanism with 
$\Delta m^2_{ei}=(0.2$--$5)\times 10^{-4}$  eV$^2$ and 
$\sin^2(2\theta)=0.6$--$1$.
\item[$\bullet$] low mass solution (LOW) via the MSW mechanism with 
$\Delta m^2_{ei}=(0.5$--$2)\times 10^{-7}$  eV$^2$ and 
$\sin^2(2\theta)=0.8$--$1$.
\end{itemize}
For oscillations into an sterile neutrino there are differences partly due to
the fact that now the survival probability depends both on the electron and
neutron density in the Sun but mainly due to the lack of neutral current
contribution to the water cerencov experiments. 
Unlike active neutrinos which lead to events in the Kamiokande and
SK detectors by interacting via neutral current with the
electrons, sterile neutrinos do not contribute to the SK
event rates.  Therefore a larger survival probability for $^8B$
neutrinos is needed to accommodate the measured rate. As a consequence
a larger contribution from $^8B$ neutrinos to the Chlorine and Gallium
experiments is expected, so that the small measured rate in Chlorine
can only be accommodated if no $^7Be$ neutrinos are present in the
flux. This is only possible in the SMA solution region, since in the
LMA and LOW regions the suppression of $^7Be$ neutrinos is not enough.
Vacuum oscillations into sterile neutrinos are also ruled out with more
than 99\% CL.

\subsection{Atmospheric Neutrinos}

Atmospheric showers are initiated when primary cosmic rays hit the
Earth's atmosphere. Secondary mesons produced in this collision,
mostly pions and kaons, decay and give rise to electron and muon
neutrino and anti-neutrinos fluxes.  There has been a
long-standing anomaly between the predicted and observed $\nu_\mu$
$/\nu_e$ ratio of the atmospheric neutrino fluxes
\cite{atmexp}. Although the absolute individual $\nu_\mu$ or $\nu_e$
fluxes are only known to within $30\%$ accuracy, different authors
agree that the $\nu_\mu$ $/\nu_e$ ratio is accurate up to a $5\%$
precision. In this resides our confidence on the atmospheric neutrino
anomaly (ANA), now strengthened by the high statistics sample
collected at the SK experiment \cite{sk99}.  
The most important feature of the atmospheric neutrino
data is that it exhibits a {\sl zenith-angle-dependent} deficit of muon 
neutrinos which is
inconsistent with expectations based on calculations of the
atmospheric neutrino fluxes. 
This experiment has marked a turning point in the significance of the
ANA. 

The most likely solution of the ANA involves neutrino
oscillations. In principle we can invoke various neutrino oscillation
channels, involving the conversion of $\nu_\mu$ into either $\nu_e$ or 
$\nu_\tau$
(active-active transitions) or the oscillation of $\nu_\mu$ into a sterile
neutrino $\nu_s$ (active-sterile transitions) \cite{atmours}. 
Oscillations into electron neutrinos are nowadays ruled out since
they cannot describe the measured angular dependence of muon-like
contained events \cite{atmours}. Moreover the most favoured range
of masses and mixings for this channel have been excluded by the 
negative results from the CHOOZ reactor experiment \cite{chooz}.

\begin{figure}
\centerline
{\protect\hbox{\psfig{file=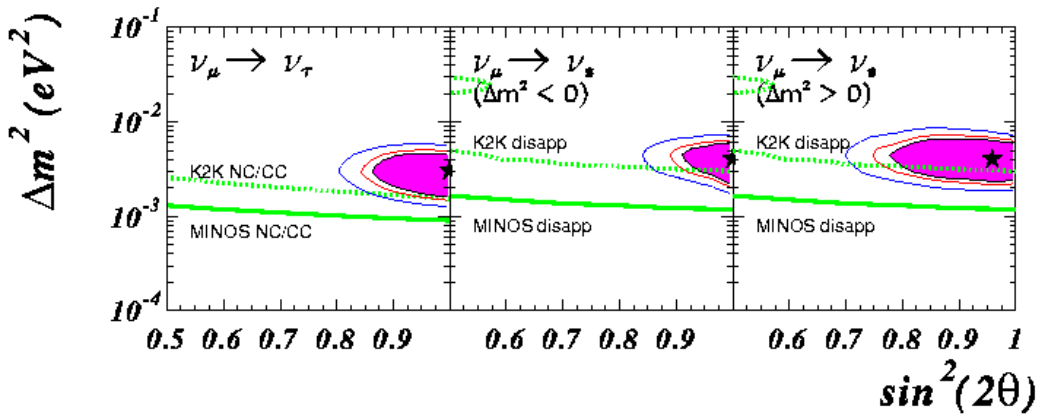,width=\textwidth}}}
\vglue -0.5cm
\caption{Allowed atmospheric oscillation parameters all for 
experiments, combined at
90 (shadowed area), 95 \% and 99 \% CL (thin solid line) for all possible
oscillation channels, from Ref.~\protect\cite{atmnew}.  
The expected sensitivity for upcoming long-baseline experiments is also 
displayed.}
\label{atmos2} 
\end{figure}

In Fig.~\ref{atmos2} we show the allowed neutrino oscillation parameters
obtained in a recent global fit of the full data set of atmospheric
neutrino data on vertex contained events at IMB, Nusex, Frejus,
Soudan, Kamiokande \cite{atmexp} and SK experiments \cite{sk99} 
as well as upward going muon data from SK, Macro and Baksan experiments
in the different oscillation channels.

The two panels corresponding to oscillations into sterile neutrinos  
in Fig.~\ref{atmos2} differ in the sign of the $\Delta m^2$ which was 
assumed in the analysis of the matter effects in the
Earth for the $\nu_\mu \to \nu_s$ oscillations. 
Notice that in all channels where matter effects play a role 
the range of acceptable $\Delta m^2$ is
shifted towards larger values, when compared with the $\nu_\mu \to
\nu_\tau$ case. This follows from looking at the relation between 
mixing {\sl in vacuo} and in matter. In fact, away from the
resonance region, independently of the sign of the matter potential,
there is a suppression of the mixing inside the Earth. As a result,
there is a lower cut in the allowed $\Delta m^2$ value, and it lies
higher than what is obtained in the data fit for the $\nu_\mu \to
\nu_\tau$ channel. 

Concerning the quality of the fits our results show 
that the best fit to the full sample is obtained for the 
$\nu_\mu \to \nu_\tau$ channel although from the global analysis
oscillations into sterile neutrinos cannot be ruled out. 
This arises mainly  from the fact that due
to matter effects the distribution for upgoing muons in the case of 
$\nu_\mu \to \nu_s$ are flatter than for $\nu_\mu \to \nu_\tau$ 
\cite{lipari}.
Data show a somehow steeper angular dependence which can be better
described by $\nu_\mu \to \nu_\tau$. This leads to the better quality
of the global fit in this channel. Pushing further this feature 
Super-Kamiokande  collaboration has presented a preliminary partial 
analysis of the angular  dependence of the through-going muon data in 
combination with the up-down asymmetry of partially contained events 
which seems to exclude the possibility $\nu_\mu \to \nu_s$ at 
the 2--$\sigma$ level \cite{sk99}. 
\subsection{LSND}
\begin{figure}
\centerline{\protect\hbox{\psfig{file=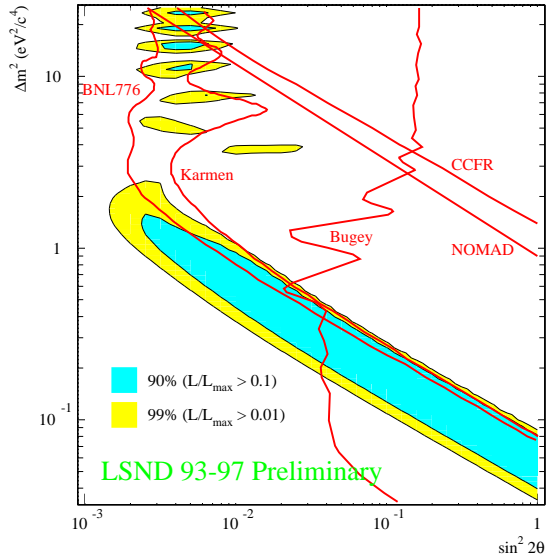,height=0.35\textheight}}}
\vglue -0.5cm
\caption{Allowed LSND oscillation parameters compared with the 90 \%
exclusion regions from other experiments.}
\label{lsnd} 
\end{figure}

Los Alamos Meson Physics Facility  (LSND) has searched  
for $\bar\nu_{\mu}\to \bar\nu_{e}$ oscillations with 
$\bar\nu_\mu$ from $\mu^+$ decay at rest \cite{lsnd}. 
The $\bar\nu_e$'s are detected in the quasi elastic process 
$\bar\nu_e\,p \to e^{+}\,n$ in correlation with a monochromatic 
photon of $2.2$  MeV arising
from the neutron capture reaction $np \to d \gamma$. In Ref.~\cite{lsnd}
they report a total of 22 events with $e^+$ energy
between 36 and $60$ MeV while  $4.6 \pm 0.6$ background
events are expected. They fit the full $e^+$ event sample in the energy
range $20 <E_e<60$ MeV by a $\chi^2$ method and the result yields
$64.3^{+18.5}_{-16.7}$ beam-related events. Subtracting the estimated
neutrino background with a correlated gamma of $12.5\pm 2.9$ events
results into an excess of 
$51.8^{+18.7}_{-16.9} \pm 8.0$ events. The interpretation of this
anomaly in terms of
$\bar \nu_\mu \to \bar \nu_e$ oscillations leads to an 
oscillation probability of ($0.31^{+0.11}_{-0.10}
\pm 0.05$)\%. 
Using a likelihood method they obtain a consistent result
of ($0.27^{+0.12}_{-0.12} \pm 0.04$)\%. 
In the two-family formalism this result  
leads to the oscillation parameters shown in
Fig.~\ref{lsnd}. The shaded regions are the 
90~\% and 99~\%  likelihood regions from LSND. Also shown are the limits from
BNL776, KARMEN1, Bugey, CCFR, and NOMAD.
\section{$\nu$--Oscillation Searches at Reactor and Accelerator Experiments} 
There are two types of laboratory experiments to search for neutrino
oscillations. In a disappearance experiment one looks for the attenuation
of a neutrino beam primarily composed of a single flavour
due to the mixing with other flavours. On the other hand in an appearance 
experiment one searches for interactions by neutrinos of a
flavour not present in the neutrino beam. Several experiments have
been searching for these signatures without any positive observation.
Their results are generally presented as exclusion areas in the two-neutrino 
oscillation  approximation. From these figures it is possible to
obtain the limits obtained by the experiments on the corresponding 
transition probabilities. This is the relevant quantity when 
interpreting the sensitivities in the framework of more--than--two--neutrino
mixing.
In table \ref{sbl} we show the limits on the different transition probabilities
from the negative results of the most restricting short baseline
experiments. 
\begin{table}[t]
\caption{90\% CL on the neutrino transition (or survival) probabilities
from the negative searches at short baseline experiments}. 
\label{sbl}
\begin{tabular}{|c|c|c|c|c|}
\hline
 Experiment & Channel & Limit (90\%) & $\Delta m^2_{min}$ (eV$^2$) & 
Reference\\ 
 \hline
  CDHSW  & $\nu_\mu\rightarrow \nu_\mu$ & $P_{\mu\mu}>0.95$ & 0.25 
&\cite{CDHSW} \\ 
  E776   & $\nu_\mu\rightarrow \nu_e$ & $P_{e\mu}<1.5\times 10^{-3}$ 
  & 0.075 &\cite{E776} \\ 
  Karmen   & $\bar\nu_\mu\rightarrow \bar\nu_e$ 
 & $P_{e\mu} <7\times 10^{-4}$ & 0.05& \cite{karmen} \\ 
  E531   & $\nu_\mu\rightarrow \nu_\tau$ & $P_{\mu\tau}<0.02$  
  & 1 & \cite{E531} \\ 
  Chorus/Nomad &  $\nu_\mu\rightarrow \nu_\tau$ & 
  $P_{\mu\tau}<6.5\times 10^{-4}$ & 0.9 &\cite{chorus}\\
 \hline
\end{tabular}
\end{table}
Due to the short path length of the neutrino in this experiments 
they are not sensitive to the low values of $\Delta m^2$ invoked
to explain both the solar and the atmospheric neutrino data. 

\begin{figure}
\centerline{\protect\hbox{\psfig{file=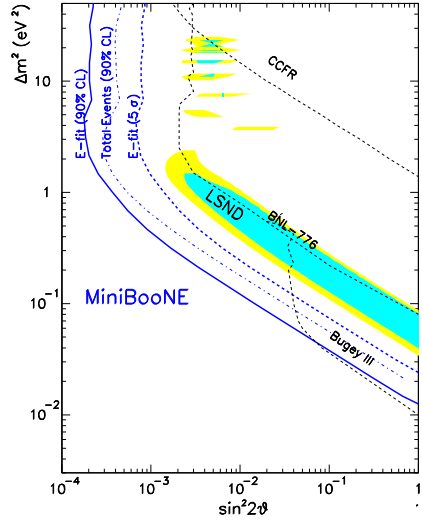,height=0.35\textheight}}}
\vglue -0.5cm
\caption{MiniBooNE 90 \% CL limits using energy--dependent fit (solid)
and total even counting (dot-dash). Also shown is the 5$\sigma$ sensitivity contour of the energy dependent fit (dashed).} 
\label{miniboone} 
\end{figure}

For the determination of the neutrino mass matrix structure the most
important short baseline  experiment will be upcoming MiniBooNE 
\cite{miniboone} experiment 
the first stage of the which is scheduled to start taking data in 2001. 
It searches for $\nu_e$ appearance in the Fermilab $\nu_\mu$ beam and it 
is specially designed to make a conclusive statement about the LSND's neutrino
oscillation evidence. In Fig. \ref{miniboone} we show the 90\%
CL limits that MiniBooNE can achieve. Should a sign be found then
the next step would be the BooNe experiment.

Smaller values of $\Delta m^2$ can be accessed at reactor experiments 
due to the lower neutrino beam energy as well as future long baseline 
experiments due to the longer distance travelled by the neutrino. In table 
\ref{lbl} we show the limits on the different 
transition probabilities from the negative results of the 
reactor experiments Bugey and CHOOZ as well as the expected sensitivities
at future long baseline experiments both at accelerator and reactors.
\begin{table}[t]
\caption{90\% CL on the neutrino transition (or survival) probabilities
from the negative searches at short baseline experiments}. 
\label{lbl}
\begin{tabular}{|c|c|c|c|c|c|}
\hline
Experiment & Type & Channel & Limit (90\%) & $\Delta m^2_{min}$ (eV$^2$) &
Ref. \\ 
\hline
 Bugey & Present Reactor & $\nu_e\rightarrow \nu_e$ & $P_{ee}>0.95$ & 
$10^{-2}$ & \cite{bugey} \\ 
CHOOZ & Present Reactor & $\nu_e\rightarrow \nu_e$ & $P_{ee}>0.91$ & 
$10^{-3}$ & \cite{chooz}\\ 
\hline
Borexino &  LBL Reactor& $\nu_e\rightarrow \nu_e$ & $P_{ee}>0.9$ 
& $10^{-6}$ & \cite{borexino} \\ 
Kamland &  LBL Reactor& $\nu_e\rightarrow \nu_e$ & $P_{ee}>0.95$ 
& $10^{-5}$ &\cite{kamland}\\ 
\hline
 K2K   &  LBL NC/CC & $\nu_\mu\rightarrow \nu_\mu$ & $P_{\mu\mu}>0.77$ 
  & $1.5\times 10^{-3}$ &\cite{k2k}\\
       &  LBL Disapp  & $\nu_\mu\rightarrow \nu_\mu$ & $P_{\mu\mu}>0.75$ 
  & $3\times 10^{-3}$ & \\
       &  LBL Appea  & $\nu_\mu\rightarrow \nu_e$ & $P_{\mu e}<0.04$ 
  & $10^{-3}$ &\\
\hline
MINOS  &   LBL NC/CC & $\nu_\mu\rightarrow \nu_\mu$ & $P_{\mu\mu}>0.99$ 
  & $9\times 10^{-4}$ &\cite{minos}\\
       &  LBL Disapp  & $\nu_\mu\rightarrow \nu_\mu$ & $P_{\mu\mu}>0.98$ 
  & $1.2\times 10^{-3}$ &\\
       &  LBL Appea  & $\nu_\mu\rightarrow \nu_e$ & 
   $P_{\mu e}<1\times 10^{-3}$  & $1 \times 10^{-3}$ &\\
 \hline
\end{tabular}
\end{table}

\section{Three--Neutrino Oscillations}

In the previous section we have discussed the evidences for neutrino
masses and mixings as usually formulated in the two neutrino oscillation
scenario. We want now to fit all the different evidences in a 
common framework and see what is our present knowledge of the neutrino
mixing and masses and how this may be improved by the upcoming 
experiments. In  doing so it is of crucial relevance the confirmation
or reprobation of the LSND result by the MiniBooNE experiment.
The three evidences can be interpreted in terms of
neutrino oscillations but with the need of three different mass
scales. Thus if the LSND is ruled out by the MiniBooNE experiment we 
could fit both solar and atmospheric data in terms of three--neutrino 
oscillations. If, on the contrary, LSND result 
stands the test of time, this would be a 
puzzling indication for the existence of a light sterile neutrino
and the need to work in a four--neutrino framework. 

In the first case, i.e. three--neutrino framework, the evolution
equation for the three neutrino flavours can be written as:
\begin{equation}
-i\frac{d\nu}{dt}=\left[U \frac{M_\nu}{2 E} U^\dagger +H_{int}\right ] \; ,
\label{evolution}
\end{equation}
where $M_\nu$ is the diagonal mass matrix for the three neutrinos and
$U$ is the unitary matrix relating the flavour and the mass basis.
$H_{int}$ is the Hamiltonian describing the neutrino interactions.
In general $U$ contains 3 mixing angles and 1 or 3 CP violating phases
depending on whether the neutrinos are Dirac or Majorana 
(For a detail discussion see Ref.\cite{Pilar}) . Here we will neglect 
the CP violating phases as they are not accessible by the existing
experiments.  
We define the unitary matrix as
\begin{equation}
U=R_{23}(\theta_{23})\times R_{13}(\theta_{13}) \times R_{12}(\theta_{12})\; ,
\end{equation}
where $R_{ij}$ is a rotation matrix in the plane $ij$.

In this framework a neutrino of definite flavour $\nu_\alpha$, 
after travelling a distance $L$ in vacuum, can be 
detected in the charged-current (CC) interaction $\nu \; N' \rightarrow  
l_\beta \; N $ with a probability
\begin{equation}
P_{\alpha\beta}=\delta_{\alpha\beta}-4\sum_{i=1}^n\sum_{j=i+1}^n
\mbox{Re}[ U_{\alpha i}U^\star_{\beta i} U^\star_{\alpha j} U_{\beta j}] 
\sin^2\left(\frac{\Delta_{ij}}{2}\right)  \; .
\end{equation}
The probability, therefore, oscillates with oscillation lengths $\Delta_{ij}$ 
given by
\begin{equation}
\frac{\Delta_{ij}}{2}=1.27 \frac{|m_i^2-m_j^2|}{\mbox{eV}^2} \frac{L/E}{\mbox{m/MeV}}=
1.27 \frac{\Delta m^2_{ij}}{\mbox{eV}^2} \frac{L/E}{\mbox{m/MeV}} \; ,
\end{equation}
where $E$ is the neutrino energy.

In general the transition probabilities will present an oscillatory
behaviour with two oscillation lengths. 
In order to explain the solar and atmospheric neutrino data we impose
the lengths to be in the range such that:
\begin{equation}
\Delta m_{12}^2\simeq \Delta m^2_{solar}< 10^{-4} \mbox{eV}^2\; , \;\;\;\;\;\;
\;\;\;\;\;\;
\Delta m_{23}^2\simeq \Delta m^2_{atm}\sim 10^{-3} \mbox{eV}^2\; .
\end{equation}
In this way, for instance, the electron and muon neutrino survival 
probabilities in vacuum are given by
\begin{eqnarray}
P_{ee}&=&1-\cos^4\theta_{13} \sin^2(2\theta_{12}) 
\sin^2\left(\frac{\Delta_{sol}}{2}\right)
 -\sin^2(2\theta_{13})
\sin^2\left(\frac{\Delta_{atm}}{2}\right) \; , \label{prob3}
 \\P_{\mu\mu}&=&1-\cos^2\theta_{13} \sin^2\theta_{23}
(1-\cos^2\theta_{13} \sin^2\theta_{23})
\sin^2\left(\frac{\Delta_{atm}}{2}\right)  \\
& &  -4(\sin\theta_{12}\cos\theta_{23} 
-\cos\theta_{12}\sin\theta_{13}\sin\theta_{23})^2 \nonumber \\
&&(\cos\theta_{12}\sin\theta_{23}
+\sin\theta_{12}\sin\theta_{13}\sin\theta_{23})^2
\sin^2\left(\frac{\Delta_{sol}}{2}\right) \; . \nonumber  
\end{eqnarray}
For the physically interesting case $\Delta m^2_{solar} \ll \Delta m^2_{atm}$ 
we find that the solar and atmospheric neutrino oscillations
decouple in the limit $\theta_{13}=0$. In this case the values of
the mixing angles $\theta_{12}$ and $\theta_{23}$ can be obtained directly
from the results of the analysis in terms of two--neutrino oscillations 
presented in the first section. Although for simplicity we have restricted 
here to vacuum oscillations, the decoupling is still valid in the presence
of matter. 

\begin{figure}
\centerline{\protect\hbox{\psfig{file=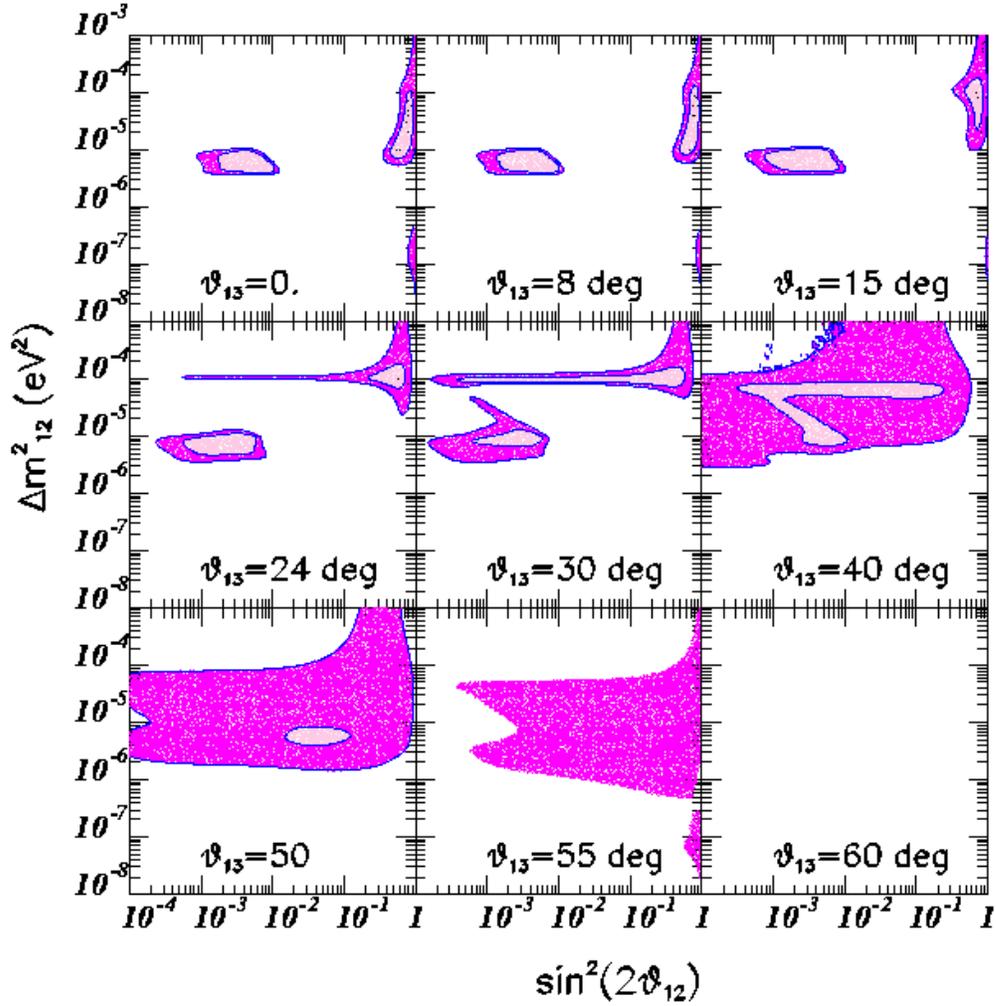,width=\textwidth}}}
\vglue -0.5cm
\caption{Allowed regions at 90\% and 99\% CL for the oscillation parameters 
$\Delta m_{12}$ and $\sin^2(\theta_{12})$ 
from the analysis of the solar neutrino 
experiments event rates in the framework of three--neutrino oscillations
for different values of the angle $\theta_{13}$.}
\label{solar3} 
\end{figure}

Deviations from the two--neutrino scenario are then determined by the
size of the mixing angle $\theta_{13}$. The first question to answer is how
the presence of this new angle affects our analysis of the solar
and atmospheric neutrino data. For vacuum solution to the solar neutrino 
problem the answer is simply given in Eq. (\ref{prob3}). 
For the case of MSW solutions it has been shown (see Ref.\cite{fogli}) 
that  
\begin{equation}
P^{3\nu}_{ee,MSW}
=\sin^4(\theta_{13})+
\cos^4(\theta_{13})P^{2\nu}_{ee,MSW} 
\end{equation}
where $P^{2\nu}_{ee,MSW}$ is obtained with the modified sun density
$N_{e}\rightarrow \cos^2(\theta_{13}) N_e $. 
In Fig. \ref{solar3} we show the allowed regions for the oscillation 
parameters $\Delta m_{12}$
and $\sin^2(\theta_{12})$ from the analysis of the solar neutrino 
experiments event rates in the framework of three--neutrino oscillations
for different values of the angle $\theta_{13}$.
As seen in the figure the effect is small unless very large values
of $\theta_{13}$ are involved. In particular for $\sin^2(\theta_{13})<0.2$
($\theta_{13}<25^\circ$) there are still two separated SMA and LMA solutions 
in the ($\Delta m^2_{12}$,$\sin^2(\theta_{12})$ ) plane (see also
Fogli's talk and reference therein).

For a detail study  of the effect of $\theta_{13}$ 
in the analysis of the atmospheric neutrino data we refer to Fogli's talk 
in these proceedings.
The conclusion is that for the presently allowed values of 
$\theta_{13}$ by the CHOOZ experiment (see below) the two--flavour analysis
of the solar and atmospheric neutrino data are good approximations in the 
determination of the allowed mass splitings 
$\Delta m_{12}^2$ and $\Delta m^2_{23}$ and mixing angles $\theta_{12}$ and 
$\theta_{23}$.
 
We must now turn to our present knowledge of the value of the mixing angle
$\theta_{13}$. Short baseline experiments cannot provide any information
on the value of this angle, as they are not sensitive to oscillations
since for both mass splitings the oscillating phase is too small. 
On the other hand experiments at reactor and long baseline experiments
can be sensitive to oscillations with $\Delta m^2_{23}$ . In Table
\ref{threefam} we show the expression for the transition probabilities
relevant for each of the experiments in the three--neutrino framework. 
\begin{table}[t]
\label{threefam}
\begin{center}
\begin{tabular}{|c|c|c|}
\hline
Experiment & Probability & Sensitivity \\
\hline
Reactor    &  $P_{ee}=1-4 s^2_{13}c^2_{13} S^2_{atm}$ & Good \\
\hline
LBL at Acc &  $P_{\mu\mu}=1-4 c^2_{13}s^2_{23} (1-c^2_{13}s^2_{23}) 
              S^2_{atm}$ & Bad \\  
     &  $P_{e\mu}=4 s^2_{13}c^2_{13} s^2_{23} S^2_{atm}$ & Good \\ 
\hline
\end{tabular}
\end{center}
\end{table}
In Table \ref{threefam} $s_i=\sin(\theta_i)$ and $c_i=\cos(\theta_i)$ 
and $S^2_{atm} =\langle \sin^2\left(\frac{\Delta_{23}}{2}\right)\rangle$.

Long baseline experiments at reactors such as Borexino and Kamland due
to the long baseline and lower reactor neutrino energy can be
sensitive to both oscillation lengths 
\begin{equation}
P_{ee}^{\mbox{LBL at reac}}=1-4 s^2_{12}c^2_{12}c_{13}^4 S^2_{sun} 
-4 s^2_{13}c^2_{13} S^2_{atm} 
\end{equation}
In this case if the solution to the solar neutrino deficit is the SMA
the contribution from the piece in $S^2_{sun}$ is very small and the 
experiments can be sensitive to $\theta_{13}$ by the observation of 
oscillations with the shorter wavelength. For the LMA solution to the
solar neutrino problem, however, both 
terms can contribute and in consequence the precision attainable on 
$\theta_{13}$ depends on the precise knowledge of the solar neutrino 
parameters $\theta_{12}$ and $\delta m^2_{12}$ which would be achieved
at future solar neutrino experiments such as SNO presently running
at Sudbury and Borexino at Gran Sasso.

\begin{figure}
\centerline{\protect\hbox{\psfig{file=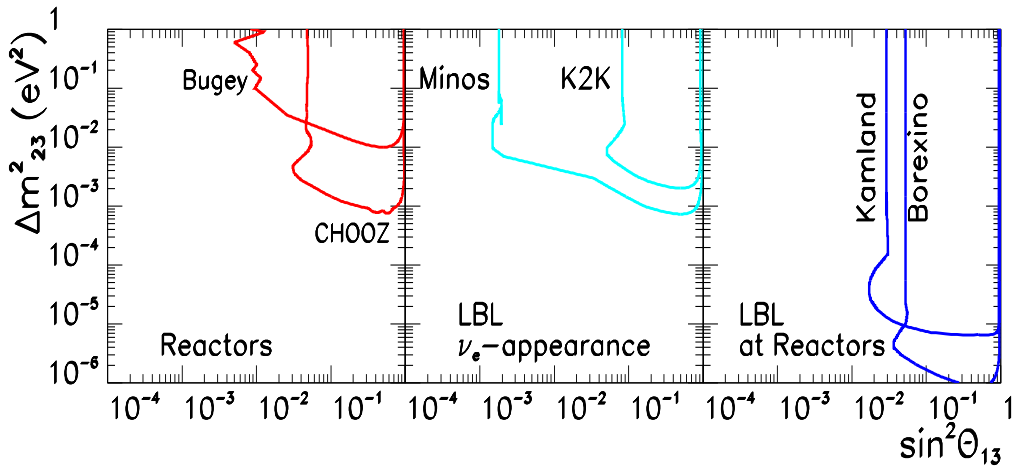,width=\textwidth,height=0.3\textheight}}}
\vglue -0.5cm
\caption{Presently excluded region in the
$\theta_{13}$ $\Delta m^2_{23}$ plane at 90\% CL by the reactor experiments as 
well as the attainable sensitivity at the future long baseline experiments.
For the central panel the region is plotted for  
the values $\sin^2(2\theta_{23})>0.8$ favoured by the atmospheric neutrinos
analysis. The regions plotted for the LBL experiments at reactors are obtained
assuming the SMA solution for the solar neutrino problem.}
\label{teta13} 
\end{figure}

In Fig. \ref{teta13} 
we plot the presently excluded region in the
$\theta_{13}$ $\Delta m^2_{23}$ plane by the present reactor experiments as 
well as the attainable sensitivity at the future long baseline experiments
listed in Table \ref{lbl}.  One must, however, take these value as the
``ultimate'' sensitivity that could be achieved at these experiments. 
The results presented in Fig. \ref{teta13}  were obtained by direct
translation of the usual two--neutrino exclusion regions to the tree--neutrino
scenario. But the original exclusion regions where obtained assuming only 
two--neutrino oscillations for the corresponding channel while in the case of 
three--neutrino mixings there may be new sources of backgrounds arising 
from other channels what can worsen the sensitivity. In order to obtain
the definite sensitivity in the angle $\theta_{13}$ the experiments should
redo their analysis in the framework of three--neutrino oscillations.

\section{Four-Neutrino Schemes}

In the previous section we have discussed the neutrino mixing parameters
assuming that the LSND result would be not confirmed by 
the MiniBooNE experiment. If the opposite holds then 
the simplest way to open the possibility of incorporating the LSND
results to the solar and atmospheric neutrino evidences is to invoke a sterile 
neutrino, i.e. one whose interaction with
standard model particles is much weaker
than the SM weak interaction so it does not affect the invisible Z decay 
width, precisely measured at LEP. The sterile neutrino must also be light
enough in order to participate in the oscillations involving the three
active neutrinos. 

\begin{figure}
\centerline{\protect\hbox{\epsfig{file=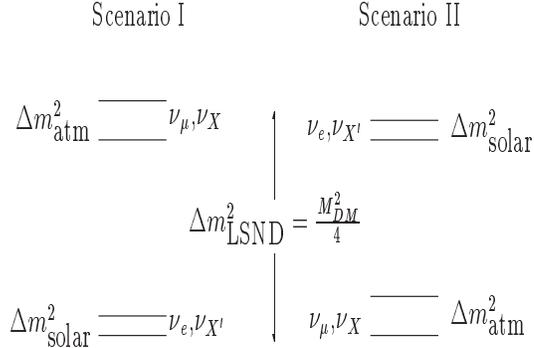,width=0.5\textwidth,height=0.2\textheight}}}
\caption{Allowed scenarios for four--neutrino oscillations.}
\label{masses} 
\end{figure}

After imposing the present constrains from the negative searches
at accelerator and reactor   neutrino oscillation experiments
one is left with two possible mass patterns as described in 
Fig.~\ref{masses} which we will call scenario I and II.
In scenario I there are two lighter neutrinos at the solar neutrino
mass scale and two maximally mixed almost degenerate eV-mass
neutrinos split by the atmospheric neutrino scale.
In scenario II the two lighter neutrinos are maximally mixed and
split by the atmospheric neutrino scale while the two
heavier neutrinos are almost degenerate separated by the solar
neutrino mass difference. In both scenarios solar neutrino data
together with reactor neutrino constrains, imply that the electron 
neutrino must be maximally projected over one of the states 
belonging to the pair split by the solar neutrino scale: the 
lighter (heavier) pair for scenario I (II). On the other hand,
atmospheric neutrino data together with the bounds from accelerator neutrino 
oscillation experiments imply that the muon neutrino must be maximally 
projected over the pair split by the atmospheric neutrino mass difference:
the heavier (lighter) pair for scenario I (II).

In both scenarios there are two possible assignments for the sterile and 
tau neutrinos which we denote by .a and .b depending on whether the tau neutrino
is maximally projected over the pair responsible for the atmospheric
neutrino oscillations and the sterile neutrino is responsible for
the solar neutrino deficit ($\nu_X=\nu_\tau$ and $\nu_{X^\prime}=\nu_s$) 
or viceversa ($\nu_X=\nu_s$ and $\nu_{X^\prime}=\nu_\tau$). For a 
more detail description of these scenarios and general consequences
we refer to the talk of C. Giunti in these proceedings \cite{four}.

In the four-neutrino scenarios the evolution equation is given as
in Eq. (\ref{evolution}) but now $U$ is a $4\times 4$ unitary matrix
which contains in general 6 mixing angles and 3 or 6 CP violating
phases \cite{four}. For the sake of simplicity, in our discussion we will 
neglect the mixing angles of the sterile neutrino with the heavy
states and the CP phases. In this case we are left with four mixing angles 
which we choose to be
\begin{equation}
U=R_{23}(\theta_{24})\times R_{23}(\theta_{23})\times R_{13}(\theta_{34}) 
\times R_{12}(\theta_{12})
\end{equation}
In general the transition probabilities will present an oscillatory
behaviour with three oscillation lengths. 
In order to explain the solar and atmospheric neutrino data, and the LSND 
result we impose the osculation lengths to be in the range such that:
\begin{eqnarray}
\Delta m_{12}^2&\simeq& \Delta m^2_{solar}< 10^{-4} \mbox{eV}^2 \nonumber \\
\Delta m_{34}^2&\simeq& \Delta m^2_{atm}\sim 10^{-3} \mbox{eV}^2  \\
\Delta m_{23}^2\simeq \Delta m_{13}^2\simeq\Delta m_{14}^2\simeq \nonumber 
\Delta m_{24}^2 &\simeq& \Delta m^2_{LSND}\sim 0.1 \;\mbox{eV}^2
\end{eqnarray}
We can work out all the survival probabilities and find that 
the solar and atmospheric neutrino oscillations
decouple in the limit $\theta_{24}=\theta_{23}=0$. In this case the values of
the mixing angles $\theta_{12}$ and $\theta_{34}$ can be obtained directly
from the results of the analysis in terms of two--neutrino oscillations 
presented in the first section. 

Deviations from the two--neutrino scenario are then determined by the
size of the mixing angles $\theta_{23}$ and $\theta_{24}$. The value of
these angles is presently limited by the reactor experiments. For
the range of mass differences invoked by the LSND experiment the most
constraining experiment is Bugey.   
The relevant transition probability is the 
$\nu_e$ survival probability. For any value of the atmospheric mass 
difference this probability always verifies
\begin{equation}
P_{lim} \leq P_{ee}^{Bugey} \leq  1-
2 c_{23}^2 c_{24}^2(1- c_{23}^2 c_{24}^2) 
\;\; \Rightarrow \;\;  c_{23}^2 c_{24}^2\geq 0.992\; .
\label{bugey}
\end{equation} 
what implies that 
\begin{equation}
\sin^2(\theta_{23})<7.5 \times 10^{-3} \;\;\;\;\;\;\;\;
\sin^2(\theta_{24})<7.5 \times 10^{-3} 
\end{equation}
or equivalently both angles must be smaller than $5^\circ$. 

Further sensitivity on the mixing angle $\theta_{23}$ is expected
at $\nu_\mu$ disappearance experiments at LBL. For instance from
the MINOS measurement of $NC/CC$ we expect 
\begin{eqnarray}
P_{lim} \leq P_{\mu\mu}^{Minos}&=&1-\sin^2(2\theta_{23}) S^2_{LSND} 
-\sin^2(2\theta_{34}) S^2_{atm} \nonumber \\
 &\leq& 1-\frac{1}{2}\sin^2(2\theta_{23}) \;\;
\Rightarrow \;\;  \sin^2(\theta_{23}) \leq 3.7 \times 10^{-3}.
\end{eqnarray}
To improve our knowledge of the mixing $\theta_{24}$ one must perform
$\nu_\tau$ appearance experiments. In these ones the relevant
survival probabilities are:
\begin{eqnarray}
P_{\mu\tau} &\simeq & \sin^2(2\theta_{23})\sin^2(\theta_{24})
S_{LSND}^2 + \sin^2(2\theta_{23})\cos^2(\theta_{24})
S_{atm}^2 \\
P_{e\tau} &\simeq &\sin^2(2\theta_{24}) S_{LSND}^2 +....
\end{eqnarray}
The presence of the shorter oscillation wavelength $S_{LSND}^2$ suggests
that in this four--neutrino scenario the best sensitivity would 
be achievable at a future high precision short baseline experiment.

\section{Conclusions}
At present, indications of non-zero neutrino masses and mixing arise
from three different sources: the solar neutrino experiments,
atmospheric neutrino data and the LSND result. The analysis of these
data in terms of two--neutrino oscillations yield three different 
oscillation scales which can only be put together in a common
framework by invoking the existence of a fourth sterile neutrino.
In this respect the most important upcoming experiment is the  
MiniBooNE experiment which searches for $\nu_e$ appearance in the Fermilab 
$\nu_\mu$ beam and it  is specially designed to make a conclusive statement 
about the LSND's neutrino oscillation evidence. 
If the LSND is ruled out by the MiniBooNE experiment we 
can fit both solar and atmospheric data in terms of three--neutrino 
oscillations. If, on the contrary, LSND result 
stands the test of time, this would be a 
puzzling indication for the existence of a light sterile neutrino
and the need to work in a four--neutrino framework. 

We have seen that in both scenarios, the existing limits on neutrino
mixings from the negative searches at reactors imply that 
the two--flavour analysis of the solar and atmospheric neutrino data are good 
approximations in the  determination of the two allowed mass splitings 
and two mixing angles.  

In the three--neutrino scenario, due to the
long oscillation lengths involved, further improvement in the additional 
mixing angle $\theta_{13}$ can 
only be achieved at long baseline experiments. In particular we have
seen that with the presently designed experiments we can  expect to reach
at most a sensitivity of about $\theta_{13} \sim 10^\circ$.

In the case of four--neutrino oscillations the presence of the shorter 
oscillation  responsible of the LSND observation  suggests
that the best sensitivity would be achievable at a future high precision 
short baseline experiment.

\noindent
{\bf Acknowledgments} \\
\noindent
We are grateful to F. Didak and J.J. Gomez-Cadenas for their kind 
hospitality in Lyon. We thank E. Akhmedov for comments. 
This work was supported by
grants DGICYT PB95-1077 and  DGICYT PB97-1261, and
by the EEC under the TMR contract ERBFMRX-CT96-0090.


\begin{thebibliography}{999}
\baselineskip=.437cm
\bibitem{solarexp}
Kamiokande Collab., Y. Fukuda {\it et al.}, Phys.
Rev. Lett. {\bf 77}, 1683(1996); Gallex Collab.,  P. Anselmann
{\it et al.}, Phys. Lett. B{\bf 342}, 440 (1995) and W. Hampel {\it et
al.}, Phys. Lett. B{\bf 388}, 364 (1996); Sage Collab.,
V. Gavrin {\it et al.}, in Neutrino {\it '96}, Proceedings of the 17th
International Conference on Neutrino Physics and Astrophysics,
Helsinki, edited by K. Huitu, K. Enqvist and J. Maalampi (World
Scientific, Singapore, 1997), p. 14; For a recent review
see the talk by T. Kirsten in  proceedings of the  
6th International Workshop on Topics in Astroparticle and 
Underground Physics, TAUP99, Paris, September 1999.
%
\bibitem{atmexp} NUSEX Collab., M. Aglietta {\sl et al.},
Europhys.  Lett.  {\bf 8}, 611 (1989); Fr\'ejus Collab., Ch.
Berger {\sl et al.}, Phys.  Lett.  {\bf B227}, 489 (1989); IMB
Collab., D. Casper {\sl et al.}, Phys. Rev. Lett.  {\bf 66},
2561 (1991); R. Becker-Szendy {\sl et al.}, Phys. Rev. {\bf D46}, 3720
(1992); Kamiokande Collab., H. S. Hirata {\sl et al.},
Phys. Lett. {\bf B205}, 416 (1988) and Phys. Lett. {\bf B280}, 146
(1992); Kamiokande Collab., Y. Fukuda {\sl et al.}, Phys.
Lett. {\bf B335}, 237 (1994); Soudan Collab., W.  W.  M Allison
{\sl et al.}, Phys.  Lett.  {\bf B391}, 491 (1997).
%
\bibitem{sk99} See talk by Y. Hayato in these proceedings
%
\bibitem{lsnd} C. Athanassopoulos, Phys.\ Rev.\ Lett.\ {\bf 75} 2650 (1995); 
Phys.\ Rev.\ Lett.\ {\bf 77} 3082 (1996); 
Phys.\ Rev.\ Lett.\ {\bf 81} 1774 (1998).
%
\bibitem{Bahcall98} J.~N. Bahcall,  Nucl.~Phys.~Proc.~Suppl.~ 
{\bf 77}, 64 (1999);
J.N.~Bahcall, M.H.~Pinsonneault, S.~Basu and
J.~Christensen-Dalsgaard, Phys.~Rev.~Lett. {\bf 78} 171 (1997);
%
\bibitem{oursun} M.C. Gonzalez-Garcia, P.C. de Holanda, C. Pe\~na-Garay and
J.W.F. Valle, hep-ph/9906469, To apear In Nucl. Phys. B.
%
\bibitem{ourfour} C. Giunti, M.C. Gonzalez-Garcia, C. Pe\~na-Garay,
hep-ph/0001101.
%
\bibitem{msw} S. P. Mikheyev and A. Yu.Smirnov, Yad.\
Fiz.\ {\bf 42}, 1441 (1985); L. Wolfenstein, Phys.\ Rev.\ {\bf D17}, 2369 
(1985).  
%
\bibitem{vacuum}
V.N.~Gribov and B.M.~Pontecorvo, Phys. Lett. {\bf 28B}, 493 (1969);
V. Barger, K. Whisnant, R.J.N. Phillips, Phys. Rev. {\bf D24}, 538
(1981); S.L.~Glashow and L.M.~Krauss, Phys. Lett. {\bf 190B}, 199
(1987); V.~Barger, R.J.~Phillips and K.~Whisnant,
Phys. Rev. Lett. {\bf 65}, 3084 (1990); S.L.~Glashow, P.J.~Kernan and
L.M.~Krauss, Phys. Lett. {\bf B445}, 412 (1999); V. Berezinsky,
G.~Fiorentini and M.~Lissia, hep-ph/9811352 and hep-ph/9904225.
%
\bibitem{atmours} 
M. C. Gonzalez-Garcia, H. Nunokawa, O. L. G. Peres, T. Stanev and
J. W. F. Valle, Phys. Rev. {\bf D58}, 033004 (1998);
M.C.~Gonzalez-Garcia, H.~Nunokawa, O.L.~Peres and J.~W.~F.~Valle,
Nucl. Phys. {\bf B543}, 3 (1999). 
%
\bibitem{atmnew} N. Fornengo, M.C. Gonzalez-Garcia, J.~W.~F.~Valle,
in preparation. 
%
\bibitem{lipari}
P. Lipari, M. Lusignoli Phys.\ Rev.\ {\bf D58}, 073005 (1998).
P. Lipari, M. Lusignoli Phys.\ Rev.\ {\bf D60}, 013003 (1999). 
%
\bibitem{chooz}
CHOOZ Collaboration, M. Apollonio {\sl et al.}. 
Phys.\ Lett.\ {\bf B420}, 397 (1998). 
%
\bibitem{CDHSW} CDHSW Collaboration, F.\ Didak {\sl et al.}, Phys.
Lett.  {\bf B134}, 281  (1984).  
%
\bibitem{E776} E776 Collaboration, L.\
Borodvsky {\sl et al.},  Phys.\ Rev.\ Lett.\ {\bf 68}, 274  (1992).  
%
\bibitem{karmen} B. Armbruster {\sl et al.}, Nucl.\ Phys.\  {\bf B38}  (Proc.
Suppl.), 235  (1995).  
%
\bibitem{E531} E531 Collaboration, Phys.\ Rev.\ Lett.\ {\bf 57}, 2898  (1986). %
\bibitem{chorus}
CHORUS Collab, , E. Eskut {\it et al.}, 
Phys.\ Lett.\ {\bf B434} (1998); NOMAD Collab. , J. Altegoer {\it et al.},
Phys.\ Lett.\ {\bf B431}, 219 (1998).
%
\bibitem{miniboone}
Andrew O. Bazarko, hep-ex/9906003. 
%
\bibitem{bugey} B.\ Achkar {\sl et al.}, Nucl.\ Phys.\ {\bf B424}, 503  (1995).
%
\bibitem{borexino} S. Schonert, 
Nucl.\ Phys.\ Proc.\ Suppl.\ {\bf 70}, 195 (1999). 
%
\bibitem{kamland}
Proposal STANFORD-HEP-98-03, July 1998.
%
\bibitem{k2k}
K. Nishikawa,  Nucl.\ Phys.\ Proc.\ Suppl.\ {\bf 77}, 198 (1999). 
%
\bibitem{minos} 
B.C. Barish, Nucl.\ Phys.\ Proc.\ Suppl.\ {\bf 70}, 227 (1999). 
%
\bibitem{Pilar} Talk by P. Hernandez in these proceedings.
%
\bibitem{fogli} See the talk of G. Fogli in these proceedings
%
\bibitem{four} See the talks of C. Giunti, A. Donini and S. Rigolini in
these proceedings.

\end{thebibliography}
\end{document}